\documentclass[12pt]{article}
\pdfoutput=1
\usepackage{latexsym,amsfonts,amsthm,amsmath,amscd,amssymb,color,mathrsfs}
\usepackage{graphicx}
\usepackage{geometry}
\usepackage[square,numbers]{natbib}
\usepackage[linktoc=page,%colorlinks,
 linkcolor=blue1,citecolor=green,urlcolor=red,runcolor=red1,linkbordercolor={0 0 1},urlbordercolor={1 0 0},citebordercolor={0 1 0},runbordercolor={1 1 1}]{hyperref}
\usepackage[utf8]{inputenc}

\newcommand\numberthis{\addtocounter{equation}{1}\tag{\theequation}}

\DeclareFontFamily{U}{mathx}{\hyphenchar\font45}
\DeclareFontShape{U}{mathx}{m}{n}{
      <5> <6> <7> <8> <9> <10>
      <10.95> <12> <14.4> <17.28> <20.74> <24.88>
      mathx10
      }{}
\DeclareSymbolFont{mathx}{U}{mathx}{m}{n}
\DeclareFontSubstitution{U}{mathx}{m}{n}
\DeclareMathAccent{\widecheck}{0}{mathx}{"71}

\makeatletter
\let\save@mathaccent\mathaccent
\newcommand*\if@single[3]{%
  \setbox0\hbox{${\mathaccent"0362{#1}}^H$}%
  \setbox2\hbox{${\mathaccent"0362{\kern0pt#1}}^H$}%
  \ifdim\ht0=\ht2 #3\else #2\fi
  }
\newcommand*\rel@kern[1]{\kern#1\dimexpr\macc@kerna}
\newcommand*\widebar[1]{\@ifnextchar^{{\wide@bar{#1}{0}}}{\wide@bar{#1}{1}}}
\newcommand*\wide@bar[2]{\if@single{#1}{\wide@bar@{#1}{#2}{1}}{\wide@bar@{#1}{#2}{2}}}
\newcommand*\wide@bar@[3]{%
  \begingroup
  \def\mathaccent##1##2{%
    \let\mathaccent\save@mathaccent
    \if#32 \let\macc@nucleus\first@char \fi
    \setbox\z@\hbox{$\macc@style{\macc@nucleus}_{}$}%
    \setbox\tw@\hbox{$\macc@style{\macc@nucleus}{}_{}$}%
    \dimen@\wd\tw@
    \advance\dimen@-\wd\z@
    \divide\dimen@ 3
    \@tempdima\wd\tw@
    \advance\@tempdima-\scriptspace
    \divide\@tempdima 10
    \advance\dimen@-\@tempdima
    \ifdim\dimen@>\z@ \dimen@0pt\fi
    \rel@kern{0.6}\kern-\dimen@
    \if#31
      \overline{\rel@kern{-0.6}\kern\dimen@\macc@nucleus\rel@kern{0.4}\kern\dimen@}%
      \advance\dimen@0.4\dimexpr\macc@kerna
      \let\final@kern#2%
      \ifdim\dimen@<\z@ \let\final@kern1\fi
      \if\final@kern1 \kern-\dimen@\fi
    \else
      \overline{\rel@kern{-0.6}\kern\dimen@#1}%
    \fi
  }%
  \macc@depth\@ne
  \let\math@bgroup\@empty \let\math@egroup\macc@set@skewchar
  \mathsurround\z@ \frozen@everymath{\mathgroup\macc@group\relax}%
  \macc@set@skewchar\relax
  \let\mathaccentV\macc@nested@a
  \if#31
    \macc@nested@a\relax111{#1}%
  \else
    \def\gobble@till@marker##1\endmarker{}%
    \futurelet\first@char\gobble@till@marker#1\endmarker
    \ifcat\noexpand\first@char A\else
      \def\first@char{}%
    \fi
    \macc@nested@a\relax111{\first@char}%
  \fi
  \endgroup
}

\def\sphere{{\mathbb S}}

\def\L{{\mathcal L}}

\def\x{{\mathbf x}}
\def\b{{\mathbf b}}
\def\y{{\mathbf y}}

\def\O{{\cal O}}
\def\F{{\cal F}}

\def\O{{\cal O}}

\newcommand{\bea}{\begin{eqnarray}}
\newcommand{\eea}{\end{eqnarray}}

\def\beq{\begin{equation}}
\def\eeq{\end{equation}}

\DeclareMathAlphabet{\mathpzc}{OT1}{pzc}{m}{it}

\begin{document}

\title{Comparing Poisson Sigma Model with A-model}
\author{F. Bonechi\footnote{\small INFN Sezione di Firenze, email: francesco.bonechi@fi.infn.it}, A.S. Cattaneo{\footnote{\small University of Zurich, email: alberto.cattaneo@math.uzh.ch}}, R. Iraso{\footnote{\small SISSA, email: riraso@sissa.it}}}

\date{}

\maketitle

\begin{abstract}
We discuss the A-model as a gauge fixing of the Poisson Sigma Model with target a symplectic structure. 
We complete the discussion in~\cite{BZ:PSM_on_sph.}, where a gauge fixing defined by a compatible complex structure was introduced, by showing how to recover the A-model hierarchy of observables in terms of the AKSZ observables. 
Moreover, we discuss the off-shell supersymmetry of the A-model as a residual BV symmetry of the gauge fixed PSM action.
\end{abstract}

\section{Introduction} 
Poisson Sigma Model (PSM) and A-model are relevant examples of two dimensional topological quantum field theories. 
The A-model is a sigma model of maps from a Riemann surface to a symplectic manifold and computes the Gromov-Witten invariants of the target manifold~\cite{witten:topological,witten:mirror}. 
The most general formulation of the A-model depends on the choice of a compatible almost complex structure~\cite{witten:topological}, but correlators are independent of it. 
When the almost complex structure is integrable, in the K\"ahler case, such a model can be obtained as a \emph{topological twist} of the supersymmetric sigma model. 
The supersymmetry transformation is responsible for the localization of the model on the space holomorphic maps and eventually for its non perturbative definition.
In the twisted formulation, this operator squares to zero only on shell, but an equivalent formulation with an off-shell cohomological supersymmetry can be obtained introducing auxiliary fields as in~\cite{witten:topological}.

The PSM is a sigma model with target a general Poisson structure, possibly degenerate; when considered on the disc it reproduces the Kontsevich formula for deformation quantization of the target Poisson structure as a correlator of boundary observables (see~\cite{CF:kontsevich_quant.}).
It is defined in terms of the AKSZ solution of the classical master equation in the Batalin-Vilkovisky formalism~\cite{AKSZ:geometry_of_BV,roytenberg:courant2}. 
It must be gauge fixed by choosing a Lagrangian submanifold of the space of fields. 
The general BV theory assures that a deformation of the Lagrangian does not affect the correlators; nevertheless, there can be different homology classes giving rise to inequivalent physical theories. 
In~\cite{BZ:PSM_on_sph.} it was shown that when one considers the PSM with target the inverse of the symplectic form of a K\"ahler manifold, the complex structure can be used to define the gauge fixing. 
The gauge fixed action coincides then, after a partial integration, with the action of the A-model. 

We complete in this letter the comparison by showing that the relevant features of the A-model, observables and supersymmetry, have a natural BV interpretation in the complex gauge fixing of the PSM.

Let us discuss first observables. 
Every de Rham cohomology class of the target manifold defines a hierarchy of observables of the A-model, whose mean values compute the Gromov-Witten invariants. 
In the AKSZ construction, there is a natural class of observables defined starting from cohomology classes of the odd vector field encoding the geometry of the target~\cite{roytenberg:courant2}. 
In the case of the PSM, this cohomology is the Lichnerowicz-Poisson (LP) cohomology; in the non degenerate case, this is canonically isomorphic to the de Rham cohomology. 
It is then natural to think that the observables of the PSM should reproduce the hierarchy of observables of the A-model after gauge fixing. 
We show that this is true but in a non trivial way. 

Indeed, for every Poisson structure the contraction with the Poisson tensor defines a map from forms to multivector fields, intertwining de Rham with LP differential.
We prove that for observables associated to multivector fields lying in the image of such a map, there is an equivalent form, up to BV operator $Q_{BV}$ and de Rham differential $\mathrm{d}$ exact terms, that in the non degenerate case and after gauge fixing reproduces the A-model hierarchy. 
We call these observables A-model like observables. 
This fact gives an interpretation of the well known independence of the Gromov-Witten invariants on the choice of the compatible complex structure in terms of independence on the choice of the gauge fixing.

Next, we discuss the residual BV symmetry. 
This is an odd symmetry of the gauge fixed action, that depends on the choice of a tubular neighbourhood of the gauge fixing Lagrangian. 
It is not true that a BV observable is closed under the residual symmetry when restricted, yet it is closed modulo equations of motion. 
Moreover, the residual symmetry squares to zero only on shell. 
We prove that in the case of the complex gauge fixing of the PSM with symplectic target, under some assumptions, there exists a choice of the tubular neighbourhood such that the residual symmetry squares to zero off shell and reproduces Witten $\mathbf{Q}$~supersymmetry with the auxiliary field considered in~\cite{witten:topological}. 
In particular, the A-model observables are closed under the residual symmetry.

In~\cite{witten:gukov} it has been discussed an approach to the quantization of symplectic manifolds based on the A-model defined on surfaces with boundary. 
This is a quantum field theoretic approach to quantization that should be compared to the results of~\cite{CF:kontsevich_quant.} and suggests a non trivial relation between the A-model and the PSM with symplectic target on surfaces with boundary that is worth investigating. 
This requires a comparison of boundary conditions of the two models that we plan to address in a future paper.

\subsubsection*{Acknowledgements}
A. S. C. acknowledges partial support of SNF Grant No. 200020-149150/1.
This research was (partly) supported by the NCCR SwissMAP, funded by the Swiss National Science Foundation, and by the COST Action MP1405 QSPACE, supported by COST (European Cooperation in Science and Technology).

\bigskip
\bigskip

\section{Residual Symmetry}\label{gauge-fixed-symm}

In this section we review the general structure of the \emph{residual symmetry} of the gauge-fixed action in BV theories.
This structure is well known (see for example~\cite{AKSZ:geometry_of_BV,schwarz:BV_geometry}), but it is useful to gather here its definition and basic properties in a convenient form for our later computations.

Recall that a classical BV theory consists of the data of a $(-1)$-symplectic manifold $(\F,\Omega)$ endowed with a cohomological Hamiltonian vector field $Q_{\mathrm{BV}}= \{S_{\mathrm{BV}},-\}$ with degree~$1$\,, where $S_{\mathrm{BV}}$ is the BV action of the theory.
Since $Q_{\mathrm{BV}}^2=0$ the BV-action is a solution of the classical master equation (CME) $\{S_{\mathrm{BV}},S_{\mathrm{BV}}\}=0$\,. If we
introduce local Darboux coordinates $\{x,x^\dagger\}$ the bracket reads
$$
\{F,G\} = \frac{\partial_r F}{\partial x^a}\frac{\partial_l G}{\partial x^\dagger_a}-\frac{\partial_r F}{\partial x^\dagger_a}\frac{\partial_l G}{\partial x^a} \;,
$$
where $\partial_r$ and $\partial_l$ denote the right and left derivative, respectively.  The CME is expressed in these local coordinates as
\beq\label{CME}
 \frac{\partial_r S_{\mathrm{BV}}}{\partial x^\dagger_a}\frac{\partial_{\ell} S_{\mathrm{BV}}}{\partial x^a}=0~.
\eeq

The gauge-fixing is performed  by restricting the action to a Lagrangian submanifold $\mathcal{L}\subset \F$\,.
The idea is that $Q_{\mathrm{BV}}$ can be projected to a vector field over $\mathcal{L}$ in such a way that the result is a symmetry of the gauge fixed BV action $S_\mathcal{L}:= S_{\mathrm{BV}}|_{\mathcal{L}}$\,.

This can be done by choosing a \emph{symplectic tubular neighbourhood} of the Lagrangian, i.e. a local symplectomorphism $\F\sim T^*[-1]\L$ restricting to the identity on~$\L$\,. 
If we denote by $\iota\colon \mathcal{L} \hookrightarrow \F$ the inclusion map and with $\pi:\F\rightarrow\L$ the projection map, the residual symmetry can be then defined by:
\beq
 Q^\pi_{\mathcal{L}} := \iota^* \circ Q_{\mathrm{BV}}\circ \pi^* ~.
\eeq

More concretely, we can think of this tubular neighbourhood as an atlas of canonical coordinates $\{x,x^\dagger\}$ adapted to $\mathcal{L}$ (i.e. $\mathcal{L}= \{x^\dagger=0\}$) such that
the transition functions between $(x,x^\dagger)$ and $(y,y^\dagger)$ are $(y=y(x),y^\dagger=(\partial x/\partial y) x^\dagger)$ so that the projection $\pi(x,x^\dagger)=x$ is well defined. 
For every function $f$ on $\mathcal{L}$ we have:
\beq
 Q^\pi_{\mathcal{L}}(f) = Q_{\mathrm{BV}}(\pi^*f) \big|_{x^\dagger=0} = -\frac{\partial_r S_{\mathrm{BV}}}{\partial x^\dagger_a}\bigg|_{x^\dagger=0}\frac{\partial_{\ell} f }{\partial x^a}~.
\eeq
In particular, it follows that $Q^\pi_\L(S_{\L})=0$ because of the CME~\eqref{CME}.

The odd version of Weinstein's theorem on the existence of a local symplectomorphism between a neighbourhood of a Lagrangian submanifold and $T^*[-1]\mathcal{L}$ was proved in~\cite{schwarz:BV_geometry}.
It must be pointed out that such a choice is non canonical and non unique: each symplectomorphism of~$\F$ into itself which keeps~$\mathcal{L}$ fixed defines a new symplectic tubular neighbourhood.
This ambiguity corresponds to the freedom to combine it with a \emph{trivial gauge transformation}:
\beq\label{res.symm.amb.}
 Q^{\pi\,a}_{\mathcal{L}}(x) \rightarrow Q^{\pi\,a}_{\mathcal{L}}(x) + \frac{\partial_r S_\mathcal{L}(x)}{\partial x^b} \mu^{ab}(x)	~,	\quad \mathrm{with} \quad	\mu^{ab} = (-1)^{(|x^a|+1)\,(|x^b|+1)}\mu^{ba}~.
\eeq
Indeed, let's see what happens if we change the tubular neighbourhood by composing the residual symmetry with a canonical transformation that leaves $\mathcal{L}$ fixed.
Let us consider a finite canonical transformation $(x,x^\dagger)\mapsto (\tilde{x},\tilde{x}^\dagger)$ generated by the function~$F(\tilde{x},x^\dagger)$~\cite{BV:1, tyutin:canonical, BV:3},
\beq
 \tilde{x}^\dagger_a=\frac{\partial F}{\partial \tilde{x}^a} \qquad x^a=\frac{\partial F}{\partial x^\dagger_a}	~.
\eeq
Since we want that the new atlas be adapted to $\mathcal{L}$, we have to impose that $\tilde{x}^\dagger = 0$ if~$x^\dagger=0$\,.
The new residual symmetry $Q^{\pi_F}_{\mathcal{L}}$ defined by~$F$ is easily found to be: %{\bf why a minus ?}:
\beq
\begin{aligned}\label{res.symm.amb.2}
 Q^{\pi_F}_{\mathcal{L}} &= -\frac{\partial_r S_{\mathrm{BV}}}{\partial \tilde{x}^\dagger_a}\bigg|_{\tilde{x}^\dagger=0} \, 
  \frac{\partial}{\partial \tilde{x}^a} = -\bigg(\frac{\partial_r S_{\mathrm{BV}}}{\partial x^\dagger_b}\,
    \frac{\partial_r x^\dagger_b}{\partial \tilde{x}^\dagger_a} + \frac{\partial_r S_{\mathrm{BV}}}{\partial x^b}\,
      \frac{\partial_r x^b}{\partial \tilde{x}^\dagger_a} \bigg)\bigg|_{\tilde{x}^\dagger=0} \frac{\partial}{\partial \tilde{x}^a}\\
 &= Q^\pi_{\mathcal{L}} - \frac{\partial_r S_\mathcal{L}}{\partial x^b}\bigg( \frac{\partial_r\partial_\ell F}{\partial x^\dagger_a \partial x^\dagger_b} \bigg)\bigg|_{x^\dagger=0}\frac{\partial}{\partial x^a}
\end{aligned}
\eeq
where we used the following relations:
$$
\frac{\partial x^b}{\partial \tilde{x}^\dagger_a}=\frac{\partial^2 F}{\partial x^\dagger_c\partial x^\dagger_b}\frac{\partial x^\dagger_c}{\partial \tilde{x}^\dagger_a}~, \qquad\qquad 
 \frac{\partial x^\dagger_c}{\partial \tilde{x}^\dagger_a}\bigg|_{x^\dagger=0}\frac{\partial}{\partial \tilde{x}^a}=\frac{\partial}{\partial x^c} ~.
$$
We see that the tensor $\mu^{ab}$ defined in~\eqref{res.symm.amb.} is, in this example:
\beq
 \mu^{ab} = \frac{\partial_r\partial_\ell F}{\partial x^\dagger_a \partial x^\dagger_b}\bigg|_{\mathcal{L}} ~,
\eeq
with the correct symmetry properties (remember that $|F|=-1$).
Notice that the variation of the residual symmetry depends only on the quadratic terms of the generating function with respect to the antifields.

The residual symmetry squares to zero only \emph{on-shell}, {\it i.e.}
\beq\label{res.symm.nilp.}
 \frac{1}{2}[Q^\pi_{\mathcal{L}},Q^\pi_{\mathcal{L}}] = \sigma^{ab} \, \frac{\partial
  S_\mathcal{L}}{\partial x^b}\frac{\partial}{\partial x^a}	~,
\eeq
where $\sigma^{ab}$ is the quadratic term in the antifield expansion of the action:
\beq\label{sviluppo_azione}
 S_{\mathrm{BV}}(x,x^\dagger) =  S_\mathcal{L}(x) -  Q^{\pi\,a}_{\mathcal{L}}(x) x^\dagger_a + 
  \frac{1}{2}x^\dagger_a\sigma^{ab}(x) x^\dagger_b + O(x^{\dagger\,3})	~.
\eeq
This allows, in particular, to define its \emph{on-shell cohomology}. 
In fact, due to the CME, the residual symmetry preserves the space of
critical points of $S_\mathcal{L}$. We call then on-shell cohomology the cohomology
of the restriction of the residual symmetry to the critical points.
Since a change of the tubular neighbourhood only modifies the residual symmetry by a trivial transformation, we have that the on-shell cohomology does not depend on the choice of tubular neighbourhood. 
We denote it as $H_{\mathrm{on}}(Q_{\mathrm{BV}},{\mathcal{L}})$\,.

The restriction of BV observables to the Lagrangian submanifold gives a map in cohomology
\beq
 H(Q_{\mathrm{BV}},\F)\rightarrow H_{\mathrm{on}}(Q_{\mathrm{BV}},{\mathcal{L}}) ~.
\eeq
Indeed the condition of being $Q_{\mathrm{BV}}$-closed, once restricted to~$\mathcal{L}$\,, reads:
\beq\label{restriction_of_observables}
 Q^\pi_\mathcal{L} (f)|_{\mathcal{L}} + V_f(S)=0 ~,
\eeq
where $V_f:=\frac{\partial_r f}{\partial x^\dagger_i}\Big|_{\mathcal{L}}\frac{\partial}{\partial x^i} \in \mathfrak{X}(\mathcal{L})$\,.
Therefore $f_\mathcal{L}$ is $Q^\pi_\mathcal{L}$-closed modulo equations of motion.
Moreover, if the symmetries of the gauge fixed action are only trivial, this map is an isomorphism (see~\cite{teitelboim:quantization},~{Thm.~18.5}).

We will be interested in the off shell residual symmetry. 
The freedom of changing the symplectic tubular neighbourhood can be used to look for a residual symmetry squaring to zero on all $\L$, not only on shell.
From~\eqref{sviluppo_azione}, a tubular neighbourhood defines the quadratic part of the BV action $\sigma=\frac{1}{2}x^\dagger_a \sigma^{ab}x^\dagger_b\in C_{-2}(T^*[-1]\L)$\,, where the grading is the (opposite) fibre degree. 
By looking at~\eqref{res.symm.nilp.}, we see that the residual symmetry $Q^\pi_\L$ squares to zero iff $\delta_{S_\L}(\sigma)=0$\,, where $\delta_{S_\L}=\iota_{\mathrm{d} S_\L}$.

When this happens, the off-shell cohomology is also defined, namely the cohomology of the residual symmetry. 
It is clear from~\eqref{restriction_of_observables} that the restriction of a BV observable to the gauge fixing Lagrangian is not in general closed under the residual symmetry.

\bigskip
\bigskip

\section{PSM and A-model and their observables}
We review in this Section the definition and basic properties of PSM and A-model.

\medskip

\subsection{A-model}
Let us introduce first the A-model following~\cite{witten:topological}. 
It is a sigma model of maps from a Riemann surface~$\Sigma$\,, with complex structure~$\varepsilon$\,, to a smooth 
$2n$-dimensional K\"ahler manifold~$M$, with complex structure~$J$\,. 
Let us introduce local coordinates $\{\sigma^\alpha\}$ on $\Sigma$ and $\{u^\mu\}$ on~$M$\,. 
Indices are raised and lowered using the K\"ahler metric.

The field content of the theory is given by a bosonic map $\phi:\Sigma\rightarrow M$ with charge $0$; a section~$\chi$ of $\phi^*(TM)$ with charge~$1$ and fermionic statistic; a one-form~$\rho$ on~$\Sigma$ with values in~$\phi^*(TM)$\,, with charge~$-1$ and fermionic statistics and a one-form~$H$ on~$\Sigma$ with values in $\phi^*(TM)$\,, with charge~$0$ and bosonic statistics. 
Finally, both~$\rho$ and~$H$ satisfy the self duality property:
\beq
  \rho^{\alpha \mu} = \varepsilon^\alpha_{\phantom{\alpha}\beta} J^\mu_{\phantom{\mu}\nu} \rho^{\beta \nu}~;	
  \qquad H^{\alpha \mu} = \varepsilon^\alpha_{\phantom{\alpha}\beta} J^\mu_{\phantom{\nu}\nu} H^{\beta \nu}~.
\eeq

The action is given by
\begin{equation}\label{action_a_model}
  S_{\varepsilon,J} = \int_\Sigma \mathrm{d}^2\sigma \Big( -\frac{1}{4}H^{\alpha \mu}H_{\alpha \mu} + H^\alpha_{\phantom{\alpha}\mu} 
  \partial_\alpha u^\mu -\mathrm{i} \rho^\alpha_{\phantom{\alpha}\mu} D_\alpha \chi^\mu -\frac{1}{8}\rho^\alpha_{\phantom{\alpha}\mu} 
  \rho_{\alpha \nu} \chi^\rho \chi^\sigma R_{\rho\sigma}^{\phantom{\mu\nu}\mu\nu} \Big) ~,
\end{equation}
where $D_\alpha \chi^\mu := \partial_\alpha \chi^\mu + \Gamma^\mu_{\nu\sigma}\chi^\nu\partial_\alpha u^\sigma$ denotes the covariant derivative with respect to the Levi Civita connection (with Christoffel symbols $\Gamma^\mu_{\nu \sigma}$) induced by the K\"ahler metric and~$R$ is the corresponding Riemann tensor.
The action is invariant under the action of the supersymmetry~$\mathbf{Q}$\,:
\beq\label{supersymmetry_a_model}
 \begin{aligned}
  &\mathbf{Q} u^\mu = \mathrm{i} \chi^\mu  ~,\\
  &\mathbf{Q} \chi^\mu = 0  ~,\\
  &\mathbf{Q} \rho_\alpha^{\phantom{\alpha}\mu}= H^{\phantom{\alpha}\mu}_{\alpha} -\mathrm{i}\Gamma^\mu_{\nu\sigma}\chi^\nu 
  \rho_\alpha^{\phantom{\alpha}\sigma} ~,\\
  &\mathbf{Q} H^{\alpha \mu} = -\frac{1}{4}\chi^\nu \chi^\sigma \Big(R^{\phantom{\nu\sigma}\mu}_{\nu\sigma\phantom{\mu}\tau} + R_{\nu\sigma\mu'\tau'}J^{\mu'\mu}
  J^{\tau'}_{\phantom{\tau'}\tau}\Big)\rho^{\alpha \tau} -\mathrm{i}\Gamma^\mu_{\nu\sigma}\chi^\nu H^{\alpha \sigma}  ~.
 \end{aligned}
\eeq
It can be seen that the odd vector field~$\mathbf{Q}$ squares to zero. 
The field~$H$ enters quadratically into the action so that it can be integrated out. 
After this integration, the action is invariant after an odd vector field that squares to zero only on shell. 
Moreover, the comparison with the PSM model is more natural including this auxiliary field, so that we will keep it without integrating.

The observables of the A-model are defined by classes of de Rham cohomology of~$M$\,. 
For an element $[\omega]\in H^k_{\mathrm{dR}}(M)$ one can define
\beq\label{A-mod_obs}
 \begin{aligned}
  &A^{(0)}_\omega = \omega_{\mu_1\ldots \mu_k} \chi^{\mu_1}\cdots \chi^{\mu_k} ~,\\
  &A^{(1)}_\omega = \mathrm{i}k\omega_{\mu_1\ldots \mu_k} \mathrm{d}u^{\mu_1}\chi^{\mu_2}\cdots \chi^{\mu_k} ~,\\  
  &A^{(2)}_\omega = -\frac{k(k-1)}{2}\omega_{\mu_1\ldots \mu_k} \mathrm{d}u^{\mu_1}\wedge\mathrm{d}u^{\mu_2}\chi^{\mu_3}\cdots \chi^{\mu_k} ~,
 \end{aligned}
\eeq
with associated A-model observables:
\beq
  A_{\omega,\gamma_k}^{(k)} = \int_{\gamma_k}A^{(k)}_\omega ~,
\eeq
where $\gamma_k$ is a $k$-cycle on $\Sigma$\,. They satisfy 
$$
\mathbf{Q} A_{\omega}^{(i)}+\mathrm{i}\mathrm{d}A_\omega^{(i-1)}=0
$$
so that $\mathbf{Q}A_{\omega,\gamma_i}^{(i)}=0$.

\medskip

\subsection{Poisson Sigma Model}
Let us introduce now the Poisson Sigma Model~(PSM). 
Let $(M,\alpha)$ be a Poisson manifold with Poisson tensor field $\alpha$ and let $\Sigma$ be a two dimensional closed surface.
The PSM in the AKSZ formalism is a two dimensional topological sigma model whose field content is the space of maps between graded manifolds $\F_\Sigma={\rm Map}(T[1]\Sigma,T^*[1]M)$\,. 
If we introduce local coordinates $x^\mu$ on~$M$ and $u^\alpha$ on~$\Sigma$\,, a point of~$\F_\Sigma$ is given by the superfields
\begin{eqnarray}
\x^\mu & = & x^\mu + \eta^{+\mu}_\alpha\theta^\alpha + \frac{1}{2} b^{+\mu}_{\alpha\beta} \theta^\alpha\theta^\beta\\
\b_\mu & = & b_\mu + \eta_{\mu\alpha}\theta^\alpha + \frac{1}{2} x^+_{\mu\alpha\beta}\theta^\alpha\theta^\beta\;, 
\end{eqnarray}
where $\theta^\alpha$ denotes the degree~$1$ coordinate of~$T[1]\Sigma$\,. 
If we change coordinates on~$M$ as $y^a=y^a(x)$, the superfields transform as:
\begin{equation}
 \label{change_of_coordinates}
 \y^a = y^a(\x) ~,\qquad \b_a = \frac{\partial x^\mu}{\partial y^a}(\x) \b_\mu ~.
\end{equation}

The space of fields $\F_\Sigma$ is a degree~$-1$ symplectic manifold with symplectic structure given by
\begin{equation}
\label{BV_symplectic_structure}
\Omega = \underset{T[1]\Sigma}{\int} \mathrm{d}u \mathrm{d}\theta \; \delta \x^\mu\wedge \delta \b_\mu ~,
\end{equation}
where $\mathrm{d}u \mathrm{d}\theta$ is the canonical Berezinian on $T[1]\Sigma$\,.
The action is given by
\begin{equation}
\label{PSM_action}
S_{\mathrm{BV}} = \underset{T[1]\Sigma}{\int} \mathrm{d}u \mathrm{d}\theta \; \Big( \mathbf{b}_\mu\mathrm{d}\mathbf{x}^\mu +\frac{1}{2}\alpha^{\mu\nu}(\mathbf{x})\mathbf{b}_\mu\mathbf{b}_\nu \Big) ~.
\end{equation}
The BV vector field $Q_{\mathrm{BV}}=\{S_{\mathrm{BV}},-\}$ reads
\beq
 \begin{aligned}\label{BV_transformation}
  &Q_{\mathrm{BV}} \mathbf{x}^\mu = \mathrm{d}\mathbf{x}^\mu + \alpha^{\mu\nu}(\mathbf{x}) \mathbf{b}_\nu ~,\\
  &Q_{\mathrm{BV}} \mathbf{b}_\mu = \mathrm{d}\mathbf{b}_\mu + \frac{1}{2}\partial_\mu\alpha^{\nu\rho}(\mathbf{x}) 
  \mathbf{b}_\nu\mathbf{b}_\rho ~,
 \end{aligned}
\eeq
where $\mathrm{d}$ is the de Rham differential on~$\Sigma$\,. 
We will be interested in the hierarchy of observables defined by Lichnerowicz-Poisson cohomology. 
We recall that the LP differential on multivector fields of $M$ is defined as $d_\alpha(v)=[\alpha,v]$, for $v\in C^\infty(T^*[1]M)\equiv\mathcal{V}^\bullet(M)$\,; it squares to zero since $\alpha$ is Poisson and we denote
by $H_{LP}(M,\alpha)$ its cohomology. 
Let $\mathrm{ev}:F_\Sigma\times T^*[1]\Sigma\rightarrow T^*[1]M$ be the evaluation map, and let us denote $\O_v=\mathrm{ev}^*(v)$\ for any 
$v\in C^\infty(T^*[1]M)$. We compute 
$$
Q_{\mathrm{BV}} (\O_v) = \mathrm{d} \O_v -\frac{1}{2} \O_{d_\alpha(v)}\;.
$$
Let us expand $\O_v=\O^{(0)}_v+\O^{(1)}_v+\O^{(2)}_v$ in form degree and assume $d_\alpha(v)=0$\,; let~$\gamma_k$ a $k$-cycle in $\Sigma$ and let $\O^{(k)}_{v,\gamma_k}\equiv\int_{\gamma_k}\O^{(k)}_v$\,, then 
$$
Q_{\mathrm{BV}}(\O^{(k)}_{v,\gamma_k})=0\;.
$$
Thus we have a hierarchy of BV observables $[\O^{(k)}_{v,\gamma_k}] \in H^\bullet(Q_{\mathrm{BV}},\F_\Sigma)$ for each $[v]\in H^\bullet_{\mathrm{LP}}(M)$\,.

Let us discuss now a subclass of these observables. 
The map
$\sharp_\alpha:\Omega^\bullet(M)\rightarrow \mathcal{V}^\bullet(M)$ defined as
\beq
  \sharp_\alpha\colon \omega_{\mu_1\ldots\mu_k} \mathrm{d}x^{\mu_1} \wedge\ldots\wedge \mathrm{d}x^{\mu_k} \mapsto  \omega_{\mu_1\ldots\mu_k}\alpha^{\mu_1\nu_1}\cdots\alpha^{\mu_k\nu_k}\partial_{\nu_1}\wedge\ldots\wedge\partial_{\nu_k}
\eeq
intertwines de Rham and LP differential $\sharp_\alpha \circ \mathrm{d} = d_\alpha \circ \sharp_\alpha$ (see~\cite{vaisman}) so that it descends to $\sharp_\alpha: H^\bullet_{\mathrm{dR}}(M)\rightarrow H^\bullet_{\mathrm{LP}}(M,\alpha)$\,. 
If the Poisson structure is non degenerate, it is an isomorphism between differential forms and multivector fields and induces an isomorphism between LP and de Rham cohomologies. 

When the LP cohomology class is in the image of this map (which is always the case when $\alpha$ is non degenerate), there is an alternative expression for the corresponding PSM observable, that we are going to discuss next. 
A long but straightforward computation shows that the PSM observable ${\cal O}_{\sharp_\alpha(\omega)}=\mathrm{ev}^*(\sharp_\alpha(\omega))$ for a closed $\omega\in\Omega^\bullet(M)$ can be written in the following form
\beq\label{sharp_obs}
 \begin{aligned}
  \mathcal{O}_{\sharp_\alpha(\omega)}^{(0)}= & 
   \frac{\mathrm{i}^k}{k!} \, \mathcal{A}_\omega^{(0)}~,	\\
  \mathcal{O}_{\sharp_\alpha(\omega)}^{(1)}= & 
   \frac{\mathrm{i}^k}{k!}\, {\mathcal A}^{(1)}_\omega +Q_{\mathrm{BV}} C^{(1)}_{\sharp_\alpha(\omega)}~, \\
  \mathcal{O}_{\sharp_\alpha(\omega)}^{(2)} = &  
   \frac{\mathrm{i}^k}{k!}\, {\mathcal A}^{(2)}_\omega + Q_{\mathrm{BV}}C^{(2)}_{\sharp_\alpha(\omega)} 
    -\mathrm{d} C^{(1)}_{\sharp_\alpha(\omega)} ~,
 \end{aligned}
\eeq
where we have defined 
 \begin{align*}
  &\mathcal{A}_\omega^{(0)} &&= && (-\mathrm{i})^k \,\omega_{\mu_1 \dots  \mu_k} b^{\mu_1}\cdots b^{\mu_k}	~,\\
  &{\mathcal A}^{(1)}_\omega &&= && \mathrm{i}k (-\mathrm{i})^{k-1} \, \omega_{\mu_1 \dots  \mu_k} \mathrm{d}x^{\mu_1}b^{\mu_2}\cdots b^{\mu_k} 	~,\\
  &{\mathcal A}^{(2)}_\omega &&= && \frac{k(k-1)}{2}(-\mathrm{i})^{k} \, \omega_{\mu_1\dots\mu_k}
   \mathrm{d}x^{\mu_1}\mathrm{d}x^{\mu_2} b^{\mu_3}\cdots  b^{\mu_k} ~,\numberthis\label{PSM_obs}\\
  &C^{(1)}_{\sharp_\alpha(\omega)} &&= &&\frac{1}{(k-1)!}\omega_{\mu_1 \dots \mu_k} \eta^{+\mu_1}b^{\mu_2}\cdots b^{\mu_k} ~,\\
  &C^{(2)}_{\sharp_\alpha(\omega)} &&= &&\omega_{\mu_1\ldots\mu_k}  \bigg(\frac{1}{(k-1)!}b^{+\mu_1}b^{\mu_2} 
   -\frac{1}{(k-2)!}\eta^{+\mu_1}\mathrm{d}x^{\mu_2} \\
 &&&&&+\frac{1}{2(k-2)!}\eta^{+\mu_1}Q_{\mathrm{BV}}\eta^{+\mu_2}\bigg)b^{\mu_3}\cdots b^{\mu_k} \\
 &&&&&+\frac{1}{2(k-1)!}\partial_\lambda 
   \omega_{\mu_1\dots\mu_k}\eta^{+\lambda}\eta^{+\mu_1}b^{\mu_2}\cdots b^{\mu_k} ~,
 \end{align*}
and~$b^\mu = \alpha^{\mu\nu}b_\nu$\,.
As a consequence of~\eqref{sharp_obs}, for each closed form $\omega$ and $k$-cycle $\gamma_k$ the observables ${\mathcal O}_{\sharp_\alpha(\omega),\gamma_k}^{(k)}$ and ${\mathcal A}_{\omega,\gamma_k}^{(k)}=\int_{\gamma_k}{\mathcal A}^{(k)}_\omega$ define the same $Q_{BV}$-cohomology class.

\bigskip
\bigskip

\section{Complex gauge fixing}
We discuss in this section how the A-model is recovered from the PSM with K\"ahler target.
Let us consider now the PSM with target the inverse of the K\"ahler form. 
In~\cite{BZ:PSM_on_sph.} a gauge fixing has been introduced such that the gauge fixed PSM action, after a partial integration, coincides with the action of the A-twist of the Supersymmetric sigma model. 

Let us introduce complex coordinates~$z$ on~$\Sigma$ and~$x^i$ on~$M$\,. 
Let us consider the Lagrangian submanifold $\L_{\varepsilon J}\subset\F_\Sigma$ defined by
$$
X^\dagger = \{x^+_i,\eta^{+ i}_z,\eta_{zi},b^{+i} ~ + {\rm c.c.}\}= 0 ~.
$$
The coordinates on~$\L_{\varepsilon J}$ are collectively called $X=\{x^{\bar{\imath}},\eta_{z\bar \imath},\eta^{+\bar \imath}_z,b_{\bar{\imath}}~+{\rm c.c.}\}$\,.
Let us consider the Christoffel symbols $\Gamma_{ij}^k$ of the Levi-Civita connection for the K\"ahler metric $\alpha^{i\bar{\jmath}}=\mathrm{i}g^{i\bar{\jmath}}$ and introduce the coordinates that 
transform tensorially:
$$
p_{\bar{z}i}=\eta_{\bar{z}i} - \Gamma_{ij}^l \eta^{+ j}_{\bar z} b_l ~.
$$
In these coordinates the gauge fixed action reads
$$
\begin{aligned}
 S_{\L_{\varepsilon J}} = \underset{\Sigma}{\int} \mathrm{d}z\mathrm{d}\bar{z}\, 
  \Big( \mathrm{i}p_{z\bar{\jmath}}\partial_{\bar{z}} x^{\bar{\jmath}} -\mathrm{i} p_{\bar{z}i} \partial_z x^i + 
   &\mathrm{i}\eta^{+i}_{\bar{z}} D_zb_i-\mathrm{i}\eta^{+\bar{\jmath}}_z D_{\bar{z}}b_{\bar{\jmath}} \\
   &+ g^{k\bar{r}}R^l_{k\bar{\jmath}i} \eta^{+i}_{\bar{z}} \eta^{+\bar{\jmath}}_z b_l b_{\bar{r}} 
    + g^{i\bar{\jmath}}p_{\bar{z}i} p_{z\bar{\jmath}} \Big)	~.
\end{aligned}    
$$
By using the transformation rules~\eqref{change_of_coordinates}, one can check that under an holomorphic change of coordinates $y^I(x^i)$ of~$M$, the corresponding transformation of fields on $\L_{\varepsilon J}$ does not depend on
momenta $X^\dagger$\,. 
The atlas $\{X,X^\dagger\}$ of adapted Darboux coordinates then fixes a symplectic tubular neighbourhood of $\L_{\varepsilon J}$ that determines the residual symmetry as 
$$
 Q_{\L_{\varepsilon J}} = b^i\frac{\delta }{\delta x^i} 
   + \Big( -\partial_{\bar{z}}x^i 
    + \Gamma^i_{kl}\eta^{+l}_{\bar{z}} b^k \Big)\frac{\delta}{\delta \eta^{+i}_{\bar{z}}} + 
 \Big( -\mathrm{i}g_{i\bar{\jmath}}D_{\bar{z}}b^{\bar{\jmath}} +  \Gamma^l_{ki}b^kp_{\bar{z}l} \Big)\frac{\delta}{\delta p_{\bar{z}i}} + \mathrm{c.c.}	~,
$$
where $b^i:= \alpha^{i\bar{\jmath}}b_{\bar{\jmath}}$\,. 
This residual BV transformation does not square to zero off shell, as one can check by a direct computation.

Let us consider a different tubular neighbourhood and look for conditions under which the corresponding residual symmetry squares to zero also off shell. 
We look for a new Darboux atlas of the space of fields adapted to the Lagrangian $\L_{\varepsilon J}$\,.
If $\widetilde{X}$ and $\widetilde{X}^\dagger$ collectively denote the new fields on $\L_{\varepsilon J}$ and their coordinate momenta respectively, then a canonical transformation can be generated by a functional $G[X,\widetilde{X}^\dagger]$\,:
\beq
 \widetilde{X}=\frac{\partial G}{\partial \widetilde{X}^\dagger}, \quad X^\dagger = \frac{\partial G}{\partial X} ~.
\eeq
This $G$ must have degree $-1$ (because $|X|+|X^\dag|=-1$), must be real and local.
Moreover, we want that $\widetilde{X}^\dagger(X,0)=0$  
and we can also ask without loss of generality that the canonical transformation is such that $\widetilde{X}(X,0)=X$\,. 
These conditions imply that there are no terms in $G$ depending only on fields and that the linear term in antifields has the form $X \widetilde{X}^\dag$\,.
This transformation will define a new tubular neighbourhood provided $\partial \widetilde{X}/\partial X^\dagger(X,0)\not=0$\,.
We will also assume, for simplicity, that the canonical transformation does not depend on any additional structure on~$\Sigma$\,.
The most general form of $G$ compatible with all the above conditions is:
\beq\label{form_G}
 G [X, \widetilde{X}^\dag] = \underset{\Sigma}{\int} \mathrm{d}z\mathrm{d}\bar{z}\, \bigg( X \widetilde{X}^\dag +\mathrm{i} \Lambda^{\bar{\imath}}_{\phantom{i}j}
 \tilde{p}_{\bar{z}\bar{\imath}}  \tilde{\eta}^{+j}_z -  \mathrm{i}\Lambda^{i}_{\phantom{i}\bar{\jmath}}\tilde{p}_{zi} \tilde{\eta}^{+\bar{\jmath}}_{\bar{z}} 
 +\mathrm{i}T_{\mu i\bar{\jmath}} b^\mu\tilde{\eta}^{+i}_z \tilde{\eta}^{+\bar{\jmath}}_{\bar{z}}\bigg)	~,
\eeq
where $\Lambda, T$ are real tensors on $M$\,.  In (\ref{form_G}) Greek letter
indices run over all coordinates, holomorphic and antiholomorphic. 
We collect here the explicit transformations 
\begin{align*}
 &\tilde{x}^\mu= x^\mu ~,				
  \quad &&\tilde{x}^+_\mu= x^+_\mu -\mathrm{i} \partial_\mu\Lambda^{\bar{\jmath}}_{\phantom{j}i} 
   \, p_{\bar{z}\bar{\jmath}} \eta^{+i}_z +\mathrm{i} \partial_\mu \Lambda^i_{\phantom{i}\bar{\jmath}} \,
    p_{zi} {\eta}^{+\bar{\jmath}}_{\bar{z}} \\
   & & & \quad\quad\quad\quad-\mathrm{i}\partial_\mu T_{\nu i\bar{\jmath}} \, 
     b^\nu \eta^{+i}_z {\eta}^{+\bar{\jmath}}_{\bar{z}} ~,\\
 &\tilde{b}^\mu= b^\mu ~,				
  \quad &&\tilde{b}^+_\mu= b^+_\mu - \mathrm{i} T_{\mu i\bar{\jmath}} \, \eta^{+i}_z {\eta}^{+\bar{\jmath}}_{\bar{z}} ~,\\
 &\tilde{p}_{\bar{z}i}= p_{\bar{z}i} + \mathrm{i}\Lambda^{\bar{\jmath}}_{\phantom{j}i} \, p_{\bar{z}\bar{\jmath}} 
  -\mathrm{i}T_{\mu i\bar{\jmath}} \, b^\mu \eta^{+\bar{\jmath}}_{\bar{z}} ~,
   \quad &&\tilde{\eta}^{+i}_{z}= {\eta}^{+i}_{z} ~,\numberthis\\
 &\tilde{\eta}^{+i}_{\bar{z}}= \eta^{+i}_{\bar{z}} 
  -\mathrm{i}\Lambda^i_{\phantom{i}\bar{\jmath}} \, \eta^{+\bar{\jmath}}_{\bar{z}} ~,		
   \quad &&\tilde{p}_{zi}= {p}_{zi} ~.
\end{align*}
We see that the new atlas is adapted to $\L_{\varepsilon J}$\,, and changes the tubular neighbourhood provided~$\Lambda$ and~$T$ are non vanishing.

One then finally computes the new residual symmetry as:
$$
\begin{aligned}
  Q^G_{\mathcal{L}_{\varepsilon J}} = Q_{\mathcal{L}_{\varepsilon J}} 
  + \Big( -\mathrm{i}g_{l\bar{\jmath}}\Lambda^{\bar{\jmath}}_{\phantom{\bar{\jmath}}i}D_{\bar{z}}b^{l} 
   +\mathrm{i} \Lambda^{\bar{\jmath}}_{\phantom{\bar{\jmath}}i} R_{s\bar{r}l\bar{\jmath}} \eta^{+l}_{\bar{z}} b^s b^{\bar{r}} 
    + b^\mu T_{\mu i \bar{\jmath}} \big(\mathrm{i}g^{l\bar{\jmath}}p_{\bar{z}l} -\partial_{\bar{z}}x^{\bar{\jmath}}\big) \Big)
     \frac{\delta}{\delta p_{\bar{z}i}} \\
  - \Lambda^{i}_{\phantom{i}\bar{\jmath}}\Big( \partial_{\bar{z}} x^{\bar{\jmath}} - \mathrm{i} g^{l\bar{\jmath}} p_{\bar{z}l} \Big) \frac{\delta}{\delta \eta^{+i}_{\bar{z}}} + 
  \mathrm{c.c.}~.
\end{aligned}
$$
It is easy to check that $(Q^G_{\mathcal{L}_{\varepsilon J}})^2$ is zero on~$x$ and~$b$\,. 
Requiring also the vanishing~of
\beq
\begin{aligned}
 (Q^G_{\mathcal{L}_{\varepsilon J}})^2\eta^{+i}_{\bar{z}} = &\big(\Lambda^{i}_{\phantom{i}\bar{\jmath}}g^{l\bar{\jmath}}\Lambda^{\bar{r}}_{\phantom{r}l}-g^{i\bar{r}}\big)\big(  
  g_{s\bar{r}} D_{\bar{z}} b^s + R_{\bar{r}k\bar{u}s} \eta^{+k}_{\bar{z}} b^s b^{\bar{u}}\big) +\\
 &+\big( \nabla_\mu \Lambda^i_{\phantom{i}\bar{r}} + \mathrm{i} g^{l\bar{\jmath}} \Lambda^i_{\phantom{i}\bar{\jmath}} 
  T_{\mu l \bar{r}} \big)\big(\mathrm{i}g^{k\bar{r}} b^\mu p_{\bar{z}k} - b^\mu \partial_{\bar{z}}x^{\bar{r}} \big)~,
\end{aligned}
\eeq
fixes the following conditions:
\beq
\begin{aligned}
 &\Lambda^{i}_{\phantom{i}\bar{\jmath}}g^{l\bar{\jmath}}\Lambda^{\bar{r}}_{\phantom{r}l}=g^{i\bar{r}} ~,\\
 &T_{\mu l\bar{r}} = -\mathrm{i}g_{i\bar{\kappa}} \Lambda^{\bar{\kappa}}_{\phantom{k}l} \nabla_\mu \Lambda^{i}_{\phantom{i}\bar{r}}~.
\end{aligned}
\eeq
It can be explicitly shown that these constraints on~$\Lambda$ and~$T$ are sufficient to have also~$(Q^G_{\mathcal{L}_{\varepsilon J}})^2 p_{\bar{z}i}=0$\,.

The possibility of choosing a tubular neighbourhood, for which the residual symmetry is cohomological, thus depends on the existence of an invertible orthogonal $(1,1)$ tensor $\Lambda$ satisfying $\Lambda J + J \Lambda=0$\,. 
There are obstructions to the existence of this structure; for instance a direct computation shows that it does not exist on $\sphere^2$\,. 
This choice is possible in many cases, for instance when $\Lambda$ is a complex structure that defines, together with $J$\,, a hyperk\"ahler structure with hyperk\"ahler metric~$g$\,. 
We remark that these data define the space filling coisotropic brane discussed in~\cite{kapustin2003remarks} and appear in the quantization scheme through the A-model described in~\cite{witten:gukov}. 
It is not clear to us if the above condition on~$\Lambda$ is also necessary and so if there is really an obstruction to the existence of the cohomological
residual symmetry.

Finally let us compare the gauge fixed action and its residual symmetry with the action of the A-model and its supersymmetry. 
Let us define
\beq
\begin{aligned}
 H_{\bar{z}}^i &:= -\partial_{\bar{z}}x^i -  \Lambda^{i}_{\phantom{i}\bar{\jmath}} (\partial_{\bar{z}}x^{\bar{\jmath}} + \alpha^{\bar{\jmath}k}p_{\bar{z}k}) ~,	
  \qquad &H_{z}^{\bar{\imath}}&=\widebar{H^i_{\phantom{z}}}}_{\!\!\!\bar{z}  ~.
\end{aligned}
\eeq
In these new variables the gauge fixed action and residual BV symmetry read
\begin{align*}
 \mathcal{S}_{\mathcal{L}\varepsilon J} = &\underset{\Sigma}{\int} \mathrm{d}z\mathrm{d}\bar{z}\, 
  \Big( -\mathrm{i}\alpha_{i\bar{\jmath}}H_{\bar{z}}^i H_{z}^{\bar{\jmath}}  -\mathrm{i} \alpha_{i\bar{\jmath}}H_{\bar{z}}^i
   \partial_zx^{\bar{\jmath}} -\mathrm{i}  \alpha_{i\bar{\jmath}}H_{{z}}^{\bar{\jmath}} \partial_{\bar{z}}x^{i} 
    +\mathrm{i}\alpha_{i\bar{\jmath}}\partial_z x^i \partial_{\bar{z}}x^{\bar{\jmath}} \\
&     - \mathrm{i}\alpha_{i\bar{\jmath}}\partial_{\bar{z}} x^i \partial_z x^{\bar{\jmath}}+  
  \mathrm{i}\alpha_{i\bar{\jmath}}\eta^{+i}_{\bar{z}} D_zb^{\bar{\jmath}} + 
  \mathrm{i}\alpha_{i\bar{\jmath}}\eta^{+\bar{\jmath}}_z D_{\bar{z}}b^i - R_{\bar{r}l\bar{\jmath}i} \eta^{+i}_{\bar{z}}
   \eta^{+\bar{\jmath}}_z b^{\bar{r}} b^l \Big)	~,\numberthis\\    
 Q^G_{\mathcal{L}_{\varepsilon J}} = &~b^i\frac{\delta }{\delta x^i} 
   + \Big( H_{\bar{z}}^i  + \Gamma^i_{kl}\eta^{+l}_{\bar{z}} b^k \Big)
    \frac{\delta}{\delta \eta^{+i}_{\bar{z}}} + \Big( - R^i_{k\bar{\jmath}l} \eta^{+l}_{\bar{z}} b^k b^{\bar{\jmath}} - 
    \Gamma^i_{kl}b^lH_{\bar{z}}^k \Big)\frac{\delta}{\delta H_{\bar{z}}^i} + \mathrm{c.c.}	
\end{align*}
which coincides, up to the topological term $\mathrm{i}\alpha_{i\bar{\jmath}}\partial_z x^i \partial_{\bar{z}}x^{\bar{\jmath}} -\mathrm{i}\alpha_{i\bar{\jmath}}\partial_{\bar{z}} x^i \partial_z x^{\bar{\jmath}}$\,, 
with the extension of the A-model with the auxiliary field $H$ given in~$(2.16)$ of~\cite{witten:topological} after the field identification $x^\mu\equiv u^\mu$, $b^\mu\equiv \mathrm{i}\chi^\mu$, $\eta^{+\bar \imath}_z \equiv \rho^{\bar \imath}_z$\,. 
With these field identifications, the restriction of the observables ${\mathcal A}_\omega^{(k)}$ defined in~\eqref{PSM_obs} on $\mathcal{L}_{\varepsilon J}$ coincides with the A-model observables $A_\omega^{(k)}$ in~\eqref{A-mod_obs}; in particular they are closed under the cohomological residual symmetry $Q^G_{\mathcal{L}_{\varepsilon J}}$\,.

\bigskip
\bigskip

\section{Conclusions}
In this Letter we compared the Poisson Sigma Model with complex gauge fixing with the A model. 
We proved that the hierarchy of observables of PSM up to~$Q_{BV}$ and~$\mathrm{d}$ exact terms coincides after complex gauge fixing with the A-model hierarchy. 
Moreover, we identified the gauge fixed action of the PSM with the action of the A-model containing the non dynamical field and we determined a symplectic tubular neighbourhood of the gauge-fixing Lagrangian such that the residual symmetry coincides with A-model supersymmetry. 
This analysis shows that the two models are the same when considered on surfaces without boundary; in particular this gives a BV explanation to the fact that Gromov-Witten invariants are independent on the choice of complex structure, 
as in the BV setting this corresponds to a choice of gauge fixing.

This analysis should be extended to the case with boundary. 
Both models provide a framework for quantization of the symplectic structure on the target. 
On the PSM side, the Kontsevich formula~\cite{konstevich:def_quant} for deformation quantization is reproduced in~\cite{CF:kontsevich_quant.} as a correlator of the model on the disk. 
On the A-model side, in~\cite{witten:gukov} the quantization is provided by the space of coisotropic branes. 
It will be natural to develop for this case the BV-BFV construction introduced in~\cite{CMRboundaries}.

\bibliography{Bibliography}
% \printbibliography
\end{document}